%% file: recsys-sampling.tex
\documentclass[sigconf,authorversion]{acmart}
\usepackage{multirow}

\def\BibTeX{{\rm B\kern-.05em{\sc i\kern-.025em b}\kern-.08emT\kern-.1667em\lower.7ex\hbox{E}\kern-.125emX}}
    
\copyrightyear{2020}
\acmYear{2020}
\setcopyright{acmlicensed}
\acmConference[CHIIR '20]{2020 Conference on Human Information Interaction and Retrieval}{March 14--18, 2020}{Vancouver, BC, Canada}
\acmBooktitle{2020 Conference on Human Information Interaction and Retrieval (CHIIR '20), March 14--18, 2020, Vancouver, BC, Canada}
\acmPrice{15.00}
\acmDOI{10.1145/3343413.3378004}
\acmISBN{978-1-4503-6892-6/20/03}

\begin{document}

%
\title{Estimating Error and Bias in Offline Evaluation Results}

%
\author{Mucun Tian}
\affiliation{%
  \institution{People \& Information Research Team\\ 
  Dept. of Computer Science,
  Boise State University}
  \city{Boise}
  \state{Idaho}
  \country{USA}
}
\email{mucuntian@u.boisestate.edu}

\author{Michael D. Ekstrand}
\orcid{0000-0003-2467-0108}
\affiliation{%
  \institution{People \& Information Research Team\\ 
  Dept. of Computer Science,
  Boise State University}
  \city{Boise}
  \state{Idaho}
  \country{USA}
}
\email{michaelekstrand@boisestate.edu}

%
\renewcommand{\shortauthors}{Tian and Ekstrand}

%
\begin{abstract}
Offline evaluations of recommender systems attempt to estimate users' satisfaction with recommendations using static data from prior user interactions.
These evaluations provide researchers and developers with first approximations of the likely performance of a new system and help weed out bad ideas before presenting them to users.
However, offline evaluation cannot accurately assess \emph{novel}, \emph{relevant} recommendations, because the most novel items were previously unknown to the user, so they are missing from the historical data and cannot be judged as relevant.

We present a simulation study to estimate the error that such missing data causes in commonly-used evaluation metrics in order to assess its prevalence and impact.
We find that missing data in the rating or observation process causes the evaluation protocol to systematically mis-estimate metric values, and in some cases erroneously determine that a popularity-based recommender outperforms even a perfect personalized recommender.
Substantial breakthroughs in recommendation quality, therefore, will be difficult to assess with existing offline techniques.
\end{abstract}


%

%

%
\maketitle
\vskip 2em

\input{introduction}
\input{background}
\input{methods}
\input{results}
\nocite{Griffiths2011}
\input{conclusion}

\begin{acks}
This work partly supported by the National Science Foundation under grant CHS 17-51278.  We thank the People and Information Research Team and Mucun's M.S. thesis committee, Sole Pera and Hoda Mehrpouyan, for their support.
\end{acks}

%

%

\bibliographystyle{ACM-Reference-Format}
\bibliography{recsys-sampling}

\appendix
\input{appendix}

\end{document}

%% file: introduction.tex
\section{Introduction}

Offline evaluation protocols for recommendation algorithms are designed to estimate how effective an algorithm is at delivering recommendations that users will find relevant.
Significant work has gone in to refining experimental protocols and metrics to better measure recommender performance, but offline evaluation still faces a fundamental difficulty: most of the data we need for a robust evaluation --- user response to various items --- is missing.
In particular, we only have data on items to which the user was somehow exposed, through an existing recommender system or other discovery mechanisms.
Offline evaluation cannot accurately measure the effectiveness of truly novel recommendations: if a recommender algorithm reliably finds items the user has never heard of, but would enjoy, the evaluation protocol will either ignore those recommendations or judge them to be irrelevant.

The existence of this problem (and related problems with missing data in recommender evaluation) is well-documented~\citep{Bellogin2011,Canamares2018,Ekstrand2017,Cremonesi2010}.
However, we do not yet understand the \emph{impact} of this missing data: how frequently, and by how much, does it lead recommender system evaluations astray?

In this paper, we present a simulation study that estimates the extent of errors caused by missing relevance data.
We simulate a data generation process consisting of complete preference data followed by observations (in the form of purchases or other consumption activities), and use these to compare evaluation results on observable data with results on complete data.
One outcome of this process is an estimate of how the offline evaluation would treat a perfect recommender, implemented as an oracle with access to the true preference data.
By varying the models and parameters in this process, we assess the sensitivity of our results to assumptions about preference and observation.
We use public data sets in multiple domains to calibrate the simulation to produce realistic observation data.
Our analysis addresses the following questions:

\begin{description}
    \item[RQ1] Which models produce more realistic simulations?
    \item[RQ2] What error does missing data cause in evaluation metrics?
    \item[RQ3] How do assumptions about data affect metric error?
    \item[RQ4] What incorrect decisions does missing data cause?
\end{description}

Our here is to model the underlying process assumed by offline evaluation and estimate its error on its own terms.
Other work will need to estimate errors resulting from those assumptions.

%% file: background.tex
\section{Related Work}
Several existing techniques attempt to measure and/or correct problems with offline evaluation.
One approach is to \emph{change the experimental protocol}.
\citet{Bellogin2012} proposed data splitting and analysis strategies to address popularity bias; these methods affect absolute metric values, but not necessarily the relative performance of algorithms \citep{Bellogin2011}.
Using random subsets of the item space as candidates for recommendation may reduce the impact of unknown relevant items~\citep{Cremonesi2010}, but it relies on unrealistically strong assumptions and likely exacerbates popularity bias \citep{Ekstrand2017}.

Another approach is to seek metrics that admit \emph{statistically unbiased estimators} with observable data.
If ratings for relevant items are missing at random, recall~\citep{Steck2010} and unnormalized DCG~\citep{Lim2015} are unbiased.
But these results limit choice of metrics and depend on assumptions unlikely to hold in actual use, as relevance is not the only influence on users' choice of items to consume or rate.

\emph{Counterfactual evaluation}~\citep{Gilotte2018, Swaminathan2015, Bottou2013} uses causal inference techniques --- often inverse propensity scoring --- to estimate how users would have responded to a different recommender algorithm.
However, it is difficult to apply to commonly-used data sets and does not yield insight into the reliability of existing evaluations.
It also cannot address the fundamental problem that concerns us in this work: if the user was never exposed to an item under the logging policy, the historical log data contains no information on its relevance.
Such items are precisely where a recommender system can produce the most benefit in many discovery-oriented applications.

\emph{Simulation} is a promising technique for studying evaluation procedures.
Simulations can produce complete ground truth and corresponding observations in a controlled manner, subject to assumptions about the structure of the data generation process.
\citet{Canamares2018} used probabilistic models to better understand the impact of popularity bias,
finding relationships between popularity bias and structural assumptions about the underlying data and inversions in the relative performance of collaborative filtering algorithms between complete and observable data in some cases.

%% file: methods.tex
\section{Simulation Methods}

\begin{figure}
\includegraphics[width=\columnwidth]{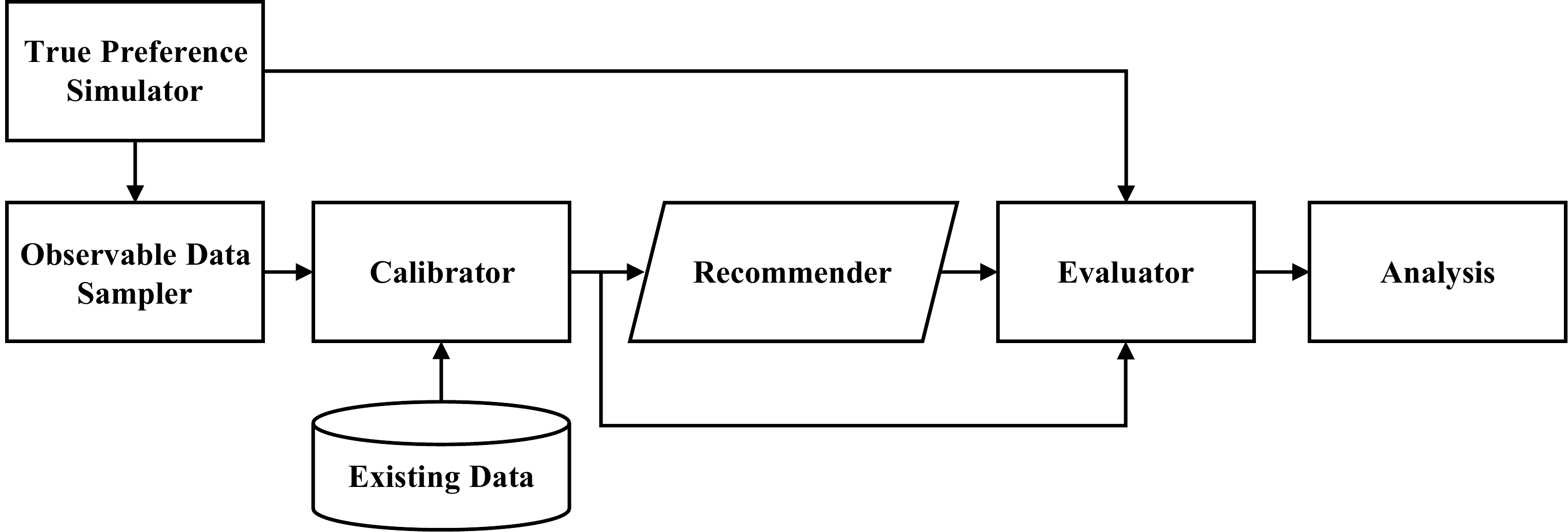}
\caption{Simulation architecture}
\label{fig:simulation-diagram}
\end{figure}

To estimate error distributions, we simulate the entire recommender system data collection and evaluation process.
Figure~\ref{fig:simulation-diagram} shows our simulation architecture; it models the user interaction that is fundamentally assumed by offline evaluation: users like items, and consume items that they (usually) like, producing records such as views or ratings that can be used to train and evaluate recommender systems.
We begin by generating (binary) user-item relevance data, then sample purchase observations from this complete truth.
We tune simulation parameters to mimic key statistics of existing public data sets, ensuring realism of simulated data.\footnote{Full experiment code is in supplementary materials. In this exposition we use the notation proposed by \citet{EKnotation}.}

We then split simulated observations into training and test data, generate recommendations for the simulated users, and measure the quality of these recommendations using both the observed test data and the underlying true preferences as ground truth.
This is the key capability our simulations enable: with access to ground truth data, because it is generated by the simulator, we can measure what the precision or reciprocal rank of a recommendation list would be if the data set were not missing data and compare it to the metric obtained from the observable data an experiment would ordinarily employ.
We can therefore compute a first approximation of the error in experimental results where we cannot access unbiased truth.
The simulated true preferences also enable us to test experimental protocols on an oracle recommender that omnisciently returns the most relevant items, irrespective of observation process.

\subsection{True Preference Models}
We use two models to simulate items being relevant to users.
Both simulation strategies produce a set $U$ of users and sets $\tilde{I}_u \subseteq I$ of items relevant to each user.
We abbreviate here for space; the appendix documents full parameterizations.
The first is preferential attachment, implemented as the three-parameter generalized Indian buffet process (\textsf{IBP}) \citep{Teh2009}.
This model is capable of producing data exhibiting power law behavior, unlike a traditional IBP.
The IBP model assumes that users like items independently; the users who like item $i$ are independent from those users who like the item $j$. This property allows us to scale up the simulation size through parallelism at the expense of realism.

The second is a correlated preference model, implemented as symmetric latent Dirichlet allocation (\textsf{LDA}) \citep{Blei2003}.
Exploiting correlations between items is fundamental to many recommendation techniques; latent feature models such as LDA provide a way to simulate such correlations.

\subsection{Observation Process}

We employ two models to produce a set of observable rated items $I_u \subseteq \tilde{I}_u$ for each user from their simulated preferences.
Both models start with $n_u$, the number of items a user will consume; since each observable user rates at least one item (or some larger number, such as 20 for MovieLens data sets) and user activity levels follow a heavy-tailed distribution, we draw $n_u$ from a truncated Pareto distribution rounded to an integer in the range $[1,|\tilde{I}_u|]$.\footnote{We also tested the truncated beta-binomial distribution; its performance is similar to the truncated Pareto. Rejection-sampling when $n_u > |\tilde{I}_u|$ produced slightly better simulations than clamping at substantial computational expense.}

\textsf{Uniform} samples $n_u$ items uniformly at random from $\tilde{I}_u$. 
This implements the assumption that observations of relevant items are \emph{missing-at-random} (MAR), so we can to compare simulation results with unbiased estimators relying on this assumption \citep{Steck2010}.

\textsf{Popular} operationalizes the idea that users are more likely to rate items that they are exposed to, and that they are more likely to be exposed to popular items.
This is one way in which observed data may violate typical MAR assumptions.
This strategy samples items with probability proportional to $|\tilde{U}_i|$ (where $\tilde{U}_i$ is the set of users who like item $i$ in the true preference data).

\begin{table*}[tb]
\caption{Simulation calibration performance, as $D_{\mathrm{KL}}(\mathrm{obs}\|\mathrm{sim})$ for each statistic, and relative loss.}
\label{tbl:calibrate}
\small
\input{calib-results.tex}
\end{table*}

\subsection{Data Sets}
\begin{table}[tb]%
\centering%
\caption{Data set summary statistics.}
\label{tbl:datasets}
\small
\begin{tabular}{lrrrr}
\toprule
\textbf{Datasets} & \textbf{Users} & \textbf{Items} & \textbf{Pairs} & \textbf{Density}\\ \midrule
ML1M & 6,040 & 3,706 & 1,000,209 & 4.47\% \\
AZM5 & 5,541 & 3,568 & 64,706 & 0.33\% \\
STMV1 & 70,912 & 10,978 & 5,094,082 & 0.65\% \\
\bottomrule
\end{tabular}
\end{table}

We set the simulation parameters by comparing their output to three data sets from different domains, summarized in Table~\ref{tbl:datasets}.
\textsf{ML1M} \citep{Harper2015-cx} contains 1M ratings of 3,706 movies from 6,040 users, where each user has at least 20 ratings.
\textsf{AZM5} \citep{He2016-um} contains 65K 5-star ratings of 3.6K digital music albums from 5.5K users; each user and item has at least 5 ratings (the ``5-core'').
\textsf{STMV1} \citep{Pathak2017} contains 5M purchases of 11K video games by 71K Australian users of the Steam game distribution service.
MovieLens and Amazon data sets are pruned~\citep{beel_data_2019}, but the Steam data is --- to our knowledge --- unpruned and accurately represents user and item distributions.

\subsection{Calibrating Simulations}

We compare synthetic data from our simulator to our reference data sets using the K-L divergence \citep{Kullback1997} between synthetic and observed distributions of four key properties: item popularity, user activity, item-item correlation, and user-user correlation.

We compute item popularity and user activity distributions by counting the frequency of each popularity or activity level (profile size) in the observed data set.
We compute item (and user) correlation distributions by sampling 1M unique item (user) pairs, with at least 5 ratings each, from the observed data and computing the cosine similarity between their ${0,1}$ rating vectors.
The 5-rating limit is to reduce the impact of uncorrelated pairs; when we did not do this, K-L divergence was dominated by the zeroes and did not meaningfully compare the distribution of nonzero correlations.

We used Gaussian process minimziation as implemented by scikit-optimize \citep{skopt} to find model parameters that minimize the K-L divergence for a particular distribution and target data set.
We optimized parameters for all four models (each combination of preference and observation model) using each distributions on each of the data sets, yielding a total of 48 tuned models.

We then retuned our models to simultaneously optimize all 4 distributions for a data set by minimizing the \emph{average relative loss} across the four distributions.
We define relative loss for a distribution by comparing the K-L divergence on that distribution to the best K-L divergence achieved on that data set with the single-objective optimizations in the previous step.
We computed mean relative loss across the four distributions to create a single integrated metric, weighting loss across all properties equally.

\subsection{Evaluation Experiments}
To simulate recommender evaluation, we held out 20\% of each user's observed items as testing data, generated 50-item recommendation lists, and computed commonly-used recommendation accuracy metrics using LensKit~\citep{Ekstrand2018}.
We computed each metric two ways: once with the held-out observable test data as ground truth, and again with the simulated true preference data. 
We repeated each experiment, including data generation, 100 times.

The \textsf{Oracle} recommender uses relevant item sets to generate perfect recommendations.
\textsf{Popular} recommends the most popular items using popularity statistics from the observable training data. 
\textsf{Random} recommends random unpurchased items for each user. 

%% file: calib-results.tex
\begin{tabular}{llrrrrrrrrr}
Data & Model  & \multicolumn{2}{c}{I-I Sim} & \multicolumn{2}{c}{Item Pop} & \multicolumn{2}{c}{U-U Sim} & \multicolumn{2}{c}{User Act} & Avg. Loss \\
\toprule
\multirow{4}{*}{ML1M} & IBP-Unif &   0.170 & 8400.00\% &    0.973 &  112.45\% &   0.227 & 3683.33\% &    0.373 &   42.37\% & 3059.54\% \\
      & IBP-Pop &   0.200 & 9900.00\% &    1.793 &  291.48\% &   0.170 & 2733.33\% &    0.388 &   48.09\% & 3243.23\% \\
      & LDA-Unif &   0.137 & 6750.00\% &    0.737 &   60.92\% &   0.082 & 1266.67\% &    0.669 &  155.34\% & 2058.23\% \\
      & LDA-Pop &   0.050 & 2400.00\% &    2.063 &  350.44\% &   0.291 & 4750.00\% &    1.084 &  313.74\% & \textbf{1953.54\%} \\
\cline{1-11}
\multirow{4}{*}{AZM5} & IBP-Unif &   0.024 &  500.00\% &    4.639 & 1007.16\% &   0.020 & 1900.00\% &    0.100 &  203.03\% &  902.55\% \\
      & IBP-Pop &   0.016 &  300.00\% &    4.170 &  895.23\% &   0.020 & 1900.00\% &    0.144 &  336.36\% &  857.90\% \\
      & LDA-Unif &   0.015 &  275.00\% &    1.029 &  145.58\% &   0.002 &  100.00\% &    0.114 &  245.45\% &  \textbf{191.51\%} \\
      & LDA-Pop &   0.035 &  775.00\% &    0.485 &   15.75\% &   0.008 &  700.00\% &    0.045 &   36.36\% &  381.78\% \\
\cline{1-11}
\multirow{4}{*}{STMV1} & IBP-Unif &   2.103 & 1402.14\% &    1.299 &  248.26\% &   1.719 & 2132.47\% &    4.503 & 1175.64\% & 1239.63\% \\
      & IBP-Pop &   0.366 &  161.43\% &    2.458 &  558.98\% &   0.770 &  900.00\% &    2.486 &  604.25\% &  \textbf{556.16\%} \\
      & LDA-Unif &   4.883 & 3387.86\% &    2.636 &  606.70\% &   6.052 & 7759.74\% &    3.100 &  778.19\% & 3133.12\% \\
      & LDA-Pop &   4.876 & 3382.86\% &    2.431 &  551.74\% &   6.041 & 7745.45\% &    3.251 &  820.96\% & 3125.25\% \\
\bottomrule
\end{tabular}

%% file: results.tex
\section{Results}
This section presents the results of running our simulations.

\subsection{Calibration Results}

Table~\ref{tbl:calibrate} shows the performance of tuned models with respect to each reference data set, averaged over 20 runs.
We show K-L divergence and relative loss for each characteristic distribution;
models with the best average loss for that data set are in bold.

\textsf{LDA}-based models fit better for ML1M and AMZM5, while the \textsf{IBP-Pop} fit best STMV1.
The high average loss on ML1M is because single-statistic optimizations were able to obtain extremely low divergence on similarity distributions (the optimal item-item K-L divergence on ML1M was 0.002), but required significant tradeoffs on similarity distributions in order to match popularity and activity level distributions.
Since the ML1M and AMZ5 are pruned, it is possible that item correlations are more important for simulating such data sets than they are for natural data sets.

\subsection{Simulation Results}

\begin{figure}[tb]
\includegraphics[width=\columnwidth]{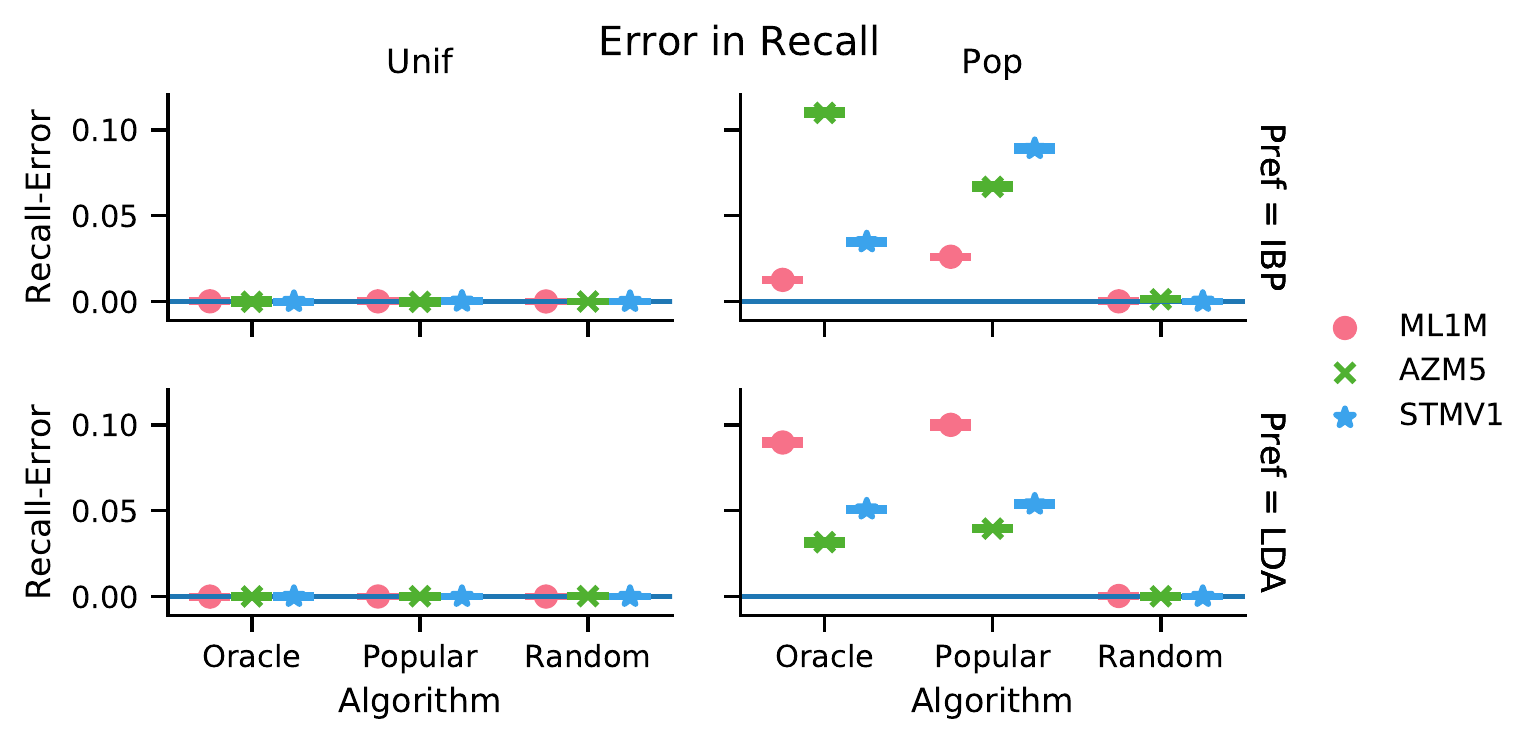}
\includegraphics[width=\columnwidth]{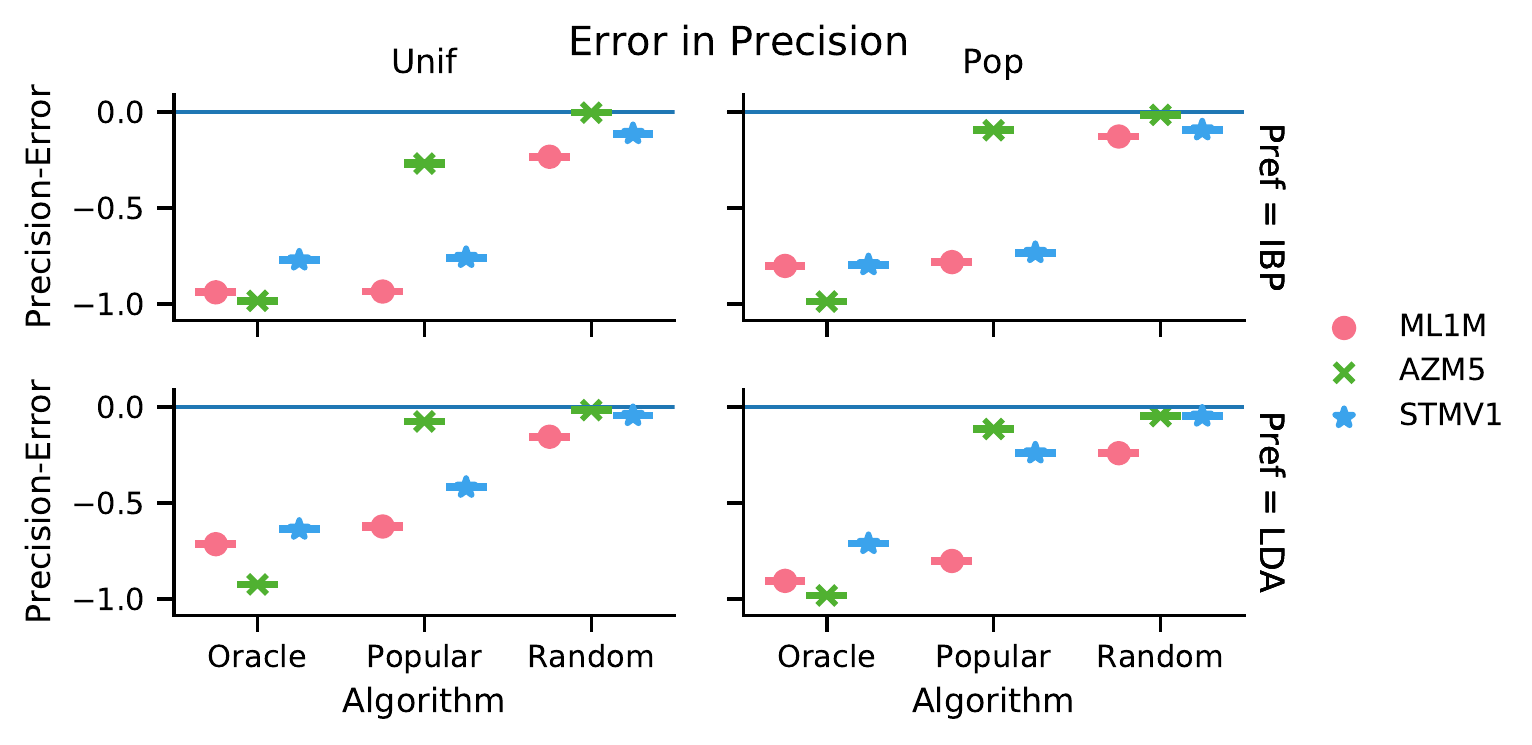}
\includegraphics[width=\columnwidth]{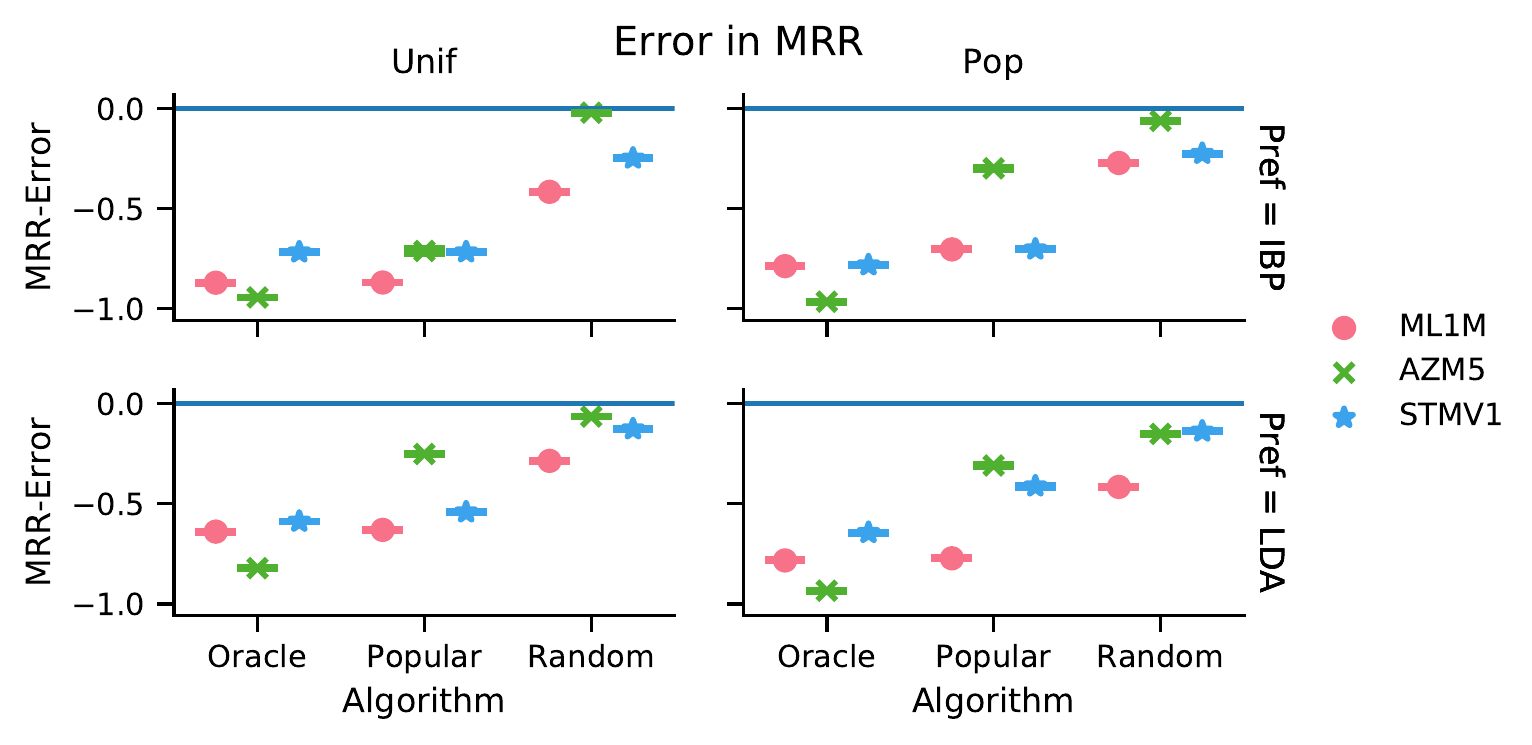}
\includegraphics[width=\columnwidth]{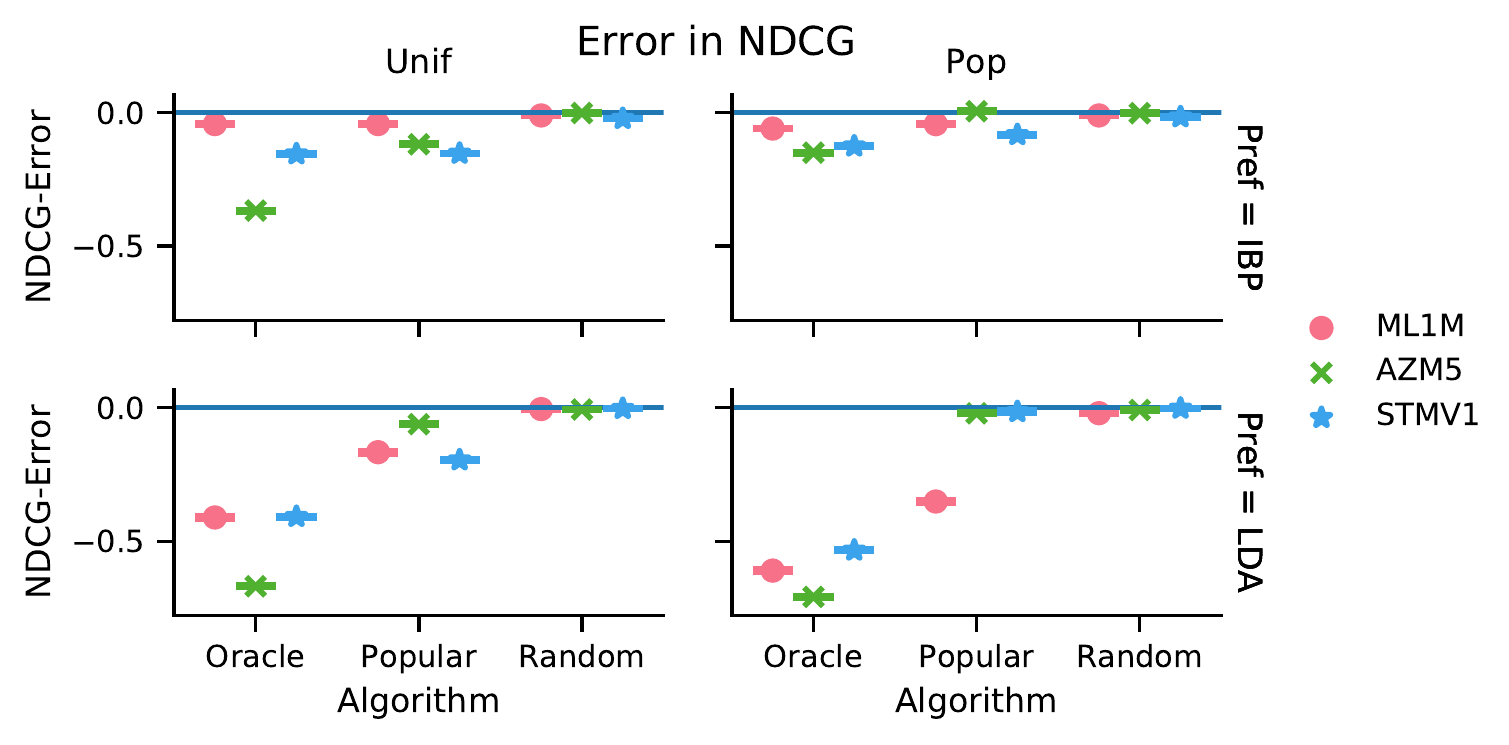}
\caption{Error of evaluation metrics ($M^{\mathrm{obs}} - M^{\mathrm{truth}}$).}
\label{fig:metric-error}
\end{figure}

Our evaluation simulations address RQ2 and RQ3.
We report the \emph{error} of each metric, defined as $M^{\mathrm{obs}} - M^{\mathrm{truth}}$ where $M^{\mathrm{obs}}$ is the metric value using observable test data as ground truth and $M^{\mathrm{truth}}$ is the value over true preference data.  Figure~\ref{fig:metric-error} shows these errors.

With the uniform observation sampler, recall has no bias (error is symmetrically distributed about 0), consistent with its status as an unbiased estimator~\citep{Steck2010}.
When observations are popularity-biased, however, observed data leads to overestimates of true recall.

Observed data generally underestimates both precision and MRR, often substantially.
Since we do not simulate users consuming irrelevant items, the set of relevant items in the true data is a superset of the observed items, increasing opportunity for recommendations to be ``correct''.
For all models, observed data underestimates these metrics more for the Oracle recommender than for Popular; this calls into question an experiment's ability to assess the performance of algorithms relative to each other, particularly when one performs extremely well.
An experiment would correctly conclude that Oracle beats Popular, but would \textit{over}-estimate the effect size.

nDCG is also biased, but its particular biases vary more between experimental conditions and recommenders.
This inconsistency suggests  it would be more difficult to correct for missing data errors in evaluations using nDCG.

\subsection{Algorithm Ranking}

\begin{table}[tb]
\caption{Percentage of runs where Oracle beats Popular.}
\label{tbl:rank-percentage}
\small
\begin{tabular}{ccccccccccc}
\multirow{1}{*}{Data} & \multirow{1}{*}{Pref} & \multirow{1}{*}{Obs} & \multicolumn{1}{c}{P@50} & \multicolumn{1}{c}{Recall} & \multicolumn{1}{c}{MRR} & \multicolumn{1}{c}{NDCG} \\ \hline
\multirow{4}{*}{ML1M} & \multirow{2}{*}{IBP} & Unif & 87 & 58 & 47 & 65 \\
 &  & Pop & 0 & 0 & 0 & 0 \\
 & \multirow{2}{*}{LDA} & Unif & 100 & 100 & 100 & 100 \\
 &  & Pop & 100 & 100 & 89 & 100 \\ \hline
\multirow{4}{*}{AZM5}
 & \multirow{2}{*}{IBP} & Unif & 100 & 100 & 100 & 100 \\
 &  & Pop & 100 & 100 & 99 & 100 \\
 & \multirow{2}{*}{LDA} & Unif & 100 & 100 & 100 & 100 \\
 &  & Pop & 100 & 100 & 100 & 100 \\ \hline
\multirow{4}{*}{STMV1} & \multirow{2}{*}{IBP} & Unif & 100 & 100 & 86 & 100 \\
 &  & Pop & 100 & 5 & 0 & 85 \\
 & \multirow{2}{*}{LDA} & Unif & 100 & 100 & 100 & 100 \\
 &  & Pop & 100 & 100 & 100 & 100  \\
 \hline
\end{tabular}
\end{table}

To address RQ4, we looked at the prevalence of \emph{rank inversions}: how often would an experiment erroneously conclude that the Popular recommender is more effective than Oracle?
Table~\ref{tbl:rank-percentage} shows these results.
While the outcomes were usually correct, in several cases they were reliably wrong.
Because this happened under IBP with Popular observations, it may be due to popularity bias severely fooling the evaluator on the observable data.
LDA did not produce this effect, indicating that extent of this error is sensitive to assumptions.

However, the most realistic scenario we examined --- the best-performing model on the unpruned Steam data set --- reliably \emph{failed} to select the oracle model over the popular one.

%% file: conclusion.tex
\section{Conclusions and Future Work}

We have simulated user preference for items and resulting consumption observations in order to estimate error and bias in the results of offline evaluations of recommender systems. To maximize realism, we calibrated our simulation models to match distributions of key characteristics of public existing data sets. 

With the exception of recall in the cases where it is already known to be an unbiased estimator, we find substantial error --- usually underestimation --- in evaluation metrics (RQ2, RQ3).
Most concerningly, we find that the \textit{degree} of error varies between algorithms in the same data and experimental condition, undermining estimates of relative differences in algorithm performance using offline evaluation protocols.
Evaluations are also sometimes fooled into misranking algorithms. This effect is sensitive to assumptions, but in what we judge to be the most realistic scenario in our study, the evaluation is unable to reliably favor an oracle recommender over a popular-item recommender on multiple metrics (RQ4).

Simulation is a promising tool for understanding recommender evaluations and characterizing their failure modes.
The simulations we present here are simple. They do not account for rating values or relative preference and do not reflect users consuming the occasional item they do not like. In future work we hope to extend this study to capture a wider array of user and algorithm behavior.

%% file: appendix.tex
\section{Model and Parameter Details}

\subsection{Preferential Attachment Preference (IBP)}

The IBP model with parameters $\alpha > 0$, $\sigma \in [0,1)$, and $c > -\sigma$ is defined as follows \citep{Teh2009}:

\begin{enumerate}
    \item The first user likes $\mathrm{Poisson}(\alpha)$ items.
    \item User $(n+1)$ likes previously-known item $i$ with probability $\frac{m_i-\sigma}{n+c}$ (where $m_i$ is the number of users who like item $i$) and likes $\mathrm{Poisson}(\alpha\frac{\Gamma(1+c)\Gamma(n+c+\sigma)}{\Gamma(n+1+c)\Gamma(c+\sigma)})$ new items.
\end{enumerate}

$c$ controls how likely the user is to rate new vs. old items.
$\sigma$ governs the power0law behavior of the generated preference matrix; $\sigma = 0$ yields a traditional IBP, with larger values yielding stronger power-law distributions of item popularity. $\alpha$ controls the density of the generated preference matrix. When $\sigma > 0$, the process generates on average $\alpha |U|^\sigma$ items; when $\sigma = 0$ and $c = 1$, it generates approximately $\alpha (\mathrm{log}|U| + \gamma)$ items on average \citep{Teh2009}, where $\gamma$ is Euler's constant \citep{Griffiths2011}.

\subsection{Correlated Preference (LDA)}

The LDA generation process \citep{Blei2003} with $K$ latent features operates as follows:

\begin{enumerate}
\item Draw $K$ feature-item vectors $\vec\phi_k \in [0,1]^{|I|}$ from $\mathrm{Dirichlet}(\beta)$.
\item For each user:
\begin{enumerate}
    \item Draw a latent feature vector $\vec\theta_u \in [0,1]^K$ from $\mathrm{Dirichlet}(\alpha)$.
    \item Draw $n_u$ (the number of items) from $\mathrm{Poisson}(\lambda)$. 
    \item Draw items $i_1, \dots, i_{n_u}$ liked by user $u$ by drawing feature $k_x \sim \mathrm{Multinomial}(\vec\theta_u)$ and $i_x$ from $\mathrm{Multinomial}(\vec\phi_{k_x})$.
\end{enumerate}
\item De-duplicate user-item pairs to produce implicit user preference samples. 
\end{enumerate}

To reduce the number of parameters for fitting efficiency, we use symmetric
LDA, where $\alpha$ is a constant vector with all values equal to $a > 0$, and likewise $\beta$ is constant $b > 0$.
These parameters $a$ and $b$ control the breadth of user preferences; when $a < 1$, the values of $\vec\theta_u$ concentrate on a few of the $K$ dimensions, making the user’s preferences concentrate on a few items if $b < 1$. The parameter $\lambda$ controls the average number of items each user likes. The parameter $K$ controls the size of the latent feature space, affecting the diversity of user-item preference patterns in the whole true preference
data.